\documentstyle[12pt]{article}

\begin{document}

\author{B.-F. Apostol, S. Stefan  \\ 
Department of Physics, University of Bucharest,\\Magurele-Bucharest, Romania
\and and M. Apostol\thanks{%
To whom correspondence should be sent}  \\ 
Department of Theoretical Physics,\\Institute of Atomic
Physics,\\Magurele-Bucharest MG-6,\\POBox MG-35,
Romania\\e-mail:apoma@theor1.ifa.ro}
\title{Entropic ''sound'' in the atmosphere }
\date{}
\maketitle

\begin{abstract}
It is shown that small, local disturbances of entropy in the atmosphere may
give rise to ''sound'' waves propagating with a velocity which depends on
the amplitude ratio of the local relative variations of temperature and
volume. This velocity is much smaller than the mean molecular velocity and
the usual, adiabatic sound velocity.
\end{abstract}
\newpage
It is well-known\cite{Palt}$^{-}$\cite{Peix} that the balance of entropy in
the atmosphere is described by the equation 
\begin{equation}
\label{1}\frac{\partial \left( \rho s\right) }{\partial t}+{\em div}\left(
\rho s{\bf c}\right) =f\;\;\;,
\end{equation}
where $\rho $ is the density of the air, $s$ is the entropy per unit mass, $%
{\bf c}$ is the transport velocity and $f$ is a source term, including the
heating rate due to the radiation, diffusion, evaporation and condensation,
as well as the internal friction. The detailed content of this source term
(which can be found, for example, in Ref.5) is not relevant for the present
discussion, since we shall focus ourselves on the homogeneous part of (\ref
{1}). Using the mass continuity the homogeneous part of the equation above
becomes 
\begin{equation}
\label{2}\frac{\partial s}{\partial t}+{\bf c}{\em grad}s=0\;\;\;,
\end{equation}
which expresses the conservation of entropy. As it is well-known,\cite{Lan}
the entropy of a closed thermodynamic system may only increase as a result
of the intermolecular collisions (the $H$ theorem). For an open system
however, as the small regions in the atmosphere, the entropy may vary
locally by the transfer of the entropy from one region to another, by the
free motion of the molecules. This variation is expressed in equation (\ref
{2}) above. This mechanism of local transfer of entropy requires the
transport velocity $\left| {\bf c}\right| $ be much smaller than the mean
molecular velocity. For a constant velocity ${\bf c}=const$ equation (\ref{2}%
) admits plane waves for the general solution. Indeed, with ${\bf c}$
oriented along the $x$-axis (\ref{2}) amounts to the wave equation 
\begin{equation}
\label{3}\frac{\partial ^2s}{\partial t^2}-c^2\frac{\partial ^2s}{\partial
x^2}=0\;\;\;,
\end{equation}
whose general solution is a superposition of plane waves with frequency $%
\omega =cq$, where $q$ is the wavevector. This suggests that small, local
perturbations of entropy may propagate in atmosphere with a velocity $c$. We
show below that these entropic waves are actually an entropic ''sound''.

\medskip\ 

Indeed, suppose that we have such an entropic wave $s\exp \left[ i\left(
qx-\omega t\right) \right] $, with the amplitude much smaller than unity; we
may also assume that locally the system is in thermodynamical equilibrium,
so that the first principle of thermodynamics (with usual notations) reads 
\begin{equation}
\label{4}dE=-pdV+Tse^{i(qx-\omega t)}\;\;.
\end{equation}
Further on, we assume an ideal gas model for the atmosphere, such that the
energy is given by $E=nc_vT+const$, and the equation of state is $pV=nT$,
where $n$ is the number of molecules per unit mass, and $c_v$ is the heat
capacity (per molecule) at constant volume. We remark that the number of
molecules per unit mass is constant, $n=1/m$, where $m$ is the mass of a
molecule. Under this assumption (\ref{4}) becomes 
\begin{equation}
\label{5}c_v\frac{dT}T+\frac{dV}V=s_0e^{i(qx-\omega t)}\;\;\;,
\end{equation}
where $s_0=s/n=sm$ is the variation of entropy per molecule. Equation (\ref
{5}) implies that the relative variations of temperature and volume may be
represented as 
\begin{equation}
\label{6}\frac{dT}T=\frac A{c_v}e^{i(qx-\omega t)}\;\;\;,
\end{equation}
\begin{equation}
\label{7}\frac{dV}V=Be^{i(qx-\omega t)}\;\;\;,
\end{equation}
with 
\begin{equation}
\label{8}A+B=s_0\;\;\;,
\end{equation}
the coefficients $A$ and $B$ being otherwise undetermined. Similarly, we get
the relative variation of pressure 
\begin{equation}
\label{9}\frac{dp}p=\left( A/c_v-B\right) e^{i(qx-\omega t)}\;\;.
\end{equation}
Equations (\ref{6}), (\ref{7}) and (\ref{9}) indicate that an entropic
''sound'', {\it i.e}. small, local variations of volume and pressure (and of
temperature, as well), is produced by the entropy disturbances, propagating
with the velocity $c$.

\medskip\ 

On the other hand, it is well-known\cite{Lamb} that such disturbances
propagate in a fluid with the velocity $v$ given by 
\begin{equation}
\label{10}v^2=1/\kappa \rho \;\;\;,
\end{equation}
where $\kappa =-(1/V)\partial V/\partial p$ is the fluid compressibility.
For $A/c_v-B\neq 0$ we get from (\ref{7}) and (\ref{9}) 
\begin{equation}
\label{11}\kappa =\frac 1p\cdot \frac B{B-A/c_v}\;\;.
\end{equation}
The (actual) sound proceeds by adiabatic, local compressions and dilations,
which correspond to putting $s_0=0$ in (\ref{8}); this leads to $A=-B$ and,
we can check that we obtain from (\ref{11}) the adiabatic compressibility $%
\kappa _{ad}=1/p\gamma $, where $\gamma =c_p/c_v$ is the adiabatic exponent
and $c_p=c_v+1$ is the heat capacity at constant pressure. We get also from (%
\ref{10}) the well-known sound velocity $c_s$ given by $c_s^2=p\gamma /\rho $%
. We can also check that we get from (\ref{11}) the isothermal
compressibility $\kappa _{is}=1/p$ for $A=0,$ which is indeed, according to (%
\ref{6}), the condition for an isothermal process. Making use of (\ref{10})
and (\ref{11}) we may express, therefore, the velocity of the entropic
''sound'' as 
\begin{equation}
\label{12}\frac p\rho \cdot \frac{B-A/c_v}B=c^2\;\;,
\end{equation}
whence one can see that the velocity $c$ is determined by the amplitude
ratio $A/B$ of the local relative variations of temperature and volume.
Assuming $c$ known equations (\ref{8}) and (\ref{12}) can be solved for the
coefficients $A$ and $B$, and we obtain 
\begin{equation}
\label{13}
\begin{array}{c}
\frac{dT}T=\frac{c_s^2-\gamma c^2}{c_s^2-c^2}\cdot \frac{s_0}{c_p}%
e^{i(qx-\omega t)}\;\;\;, \\  \\ 
\frac{dV}V=\frac{c_s^2}{c_s^2-c^2}\cdot \frac{s_0}{c_p}e^{i(qx-\omega
t)}\;\;\;, \\  \\ 
\frac{dp}p=-\frac{\gamma c^2}{c_s^2-c^2}\cdot \frac{s_0}{c_p}e^{i(qx-\omega
t)}\;\;.
\end{array}
\end{equation}
Since $c$ is much smaller than the mean molecular velocity we have also $%
c\ll c_s$, so that these perturbations do propagate indeed as an entropic
''sound'' with the velocity $c$, {\it i.e.} the entropic ''sound'' does
indeed exist. For $c$ approaching $c_s$ we see from (\ref{12}) that $B$
approaches $-A$, {\it i.e}., according to (\ref{8}), the entropy variations
vanish ($s_0\rightarrow 0$), and the entropic ''sound'' becomes the usual,
adiabatic sound.

\medskip\ 

Equations (\ref{8}) and (\ref{12}), which were solved for the coefficients $%
A $ and $B$ above, i.e. for the small amplitudes of the relative variations
of the temperature and, respectively, volume (the relative variations of the
pressure are obtained from the equation of state), may also be viewed in
another way. We may either consider that the velocity $c$ is given, and then
we obtain the amplitudes $A$ and $B$, or consider these temperature and
volume variations as being fixed (such as to satisfy the first law of
thermodynamics), by various, undetermined circumstances which caused the
initial entropy disturbance, and then we get the velocity $c$. One can see
that the velocity $c$ is therefore not fixed, but depends on the relative
magnitude of the original local disturbances of temperature and volume (or
pressure). In addition, we may also remark that if the masses of air in the
atmosphere are in motion with an additional transport velocity $u$,{\it \
i.e.} with a wind velocity $u$, then the entropy waves may be written as $%
s\exp \left[ i\left( q(x+ut)-\omega t\right) \right] $, such that the
frequency is given by $\omega =(c+u)q$, in agreement with the Galilei
principle of translational symmetry. As is well-known the wind velocity $u$
itself is usually much smaller than the mean molecular velocity (and the
adiabatic sound velocity).

\medskip\

\end{document}